\documentclass[aps,pra,twocolumn,longbibliography]{revtex4-1}
\usepackage[colorlinks=true, citecolor=red, urlcolor=blue ]{hyperref}
 \usepackage[utf8]{inputenc}
\usepackage{graphicx}   % Include figure files
\usepackage{dcolumn}    % Align table columns on decimal point
\usepackage{bm}         % Bold math
\usepackage{hyperref}   % Add hypertext capabilities
\usepackage{xcolor}     % Text coloring for revisions
\usepackage{amsmath}
\usepackage{relsize}
\usepackage{comment}
\usepackage[caption=false]{subfig} 
\usepackage{ragged2e}
\usepackage{booktabs}
\usepackage{indentfirst}
\usepackage{epsfig}
\usepackage{amsthm} % provides theorem and proof environments
 \usepackage{amssymb}
 \usepackage{placeins} 
 \usepackage{multirow}
\usepackage{subfig}   % NOT subfigure (obsolete)
\usepackage[normalem]{ulem}
\usepackage{mathrsfs}

\newcommand{\tco}{t_{c}^{(1)}}
\newcommand{\tmo}{t_{\downarrow}^{(1)}}
\newcommand{\tpo}{t_{\uparrow}^{(1)}}

\begin{document}

% Title of the paper
%\title{Scaling of temperature in quantum refrigerator with spin-star environments}

%\title{Quantum refrigerator embedded in spin stars:\\Scalings of temperature and refrigeration time with environment size}

\title{Dynamical Quantum Phase Transitions in Boundary Time Crystals}

% Authors and affiliations
\author{Sukrut Mondkar}
\email{sukrutmondkar@gmail.com}
\affiliation{Harish-Chandra Research Institute, A CI of Homi Bhabha National Institute, Chhatnag Road, Jhusi,
Prayagraj (Allahabad) 211019, India}

\author{Priya Ghosh}
\email{priyaghosh1155@gmail.com}
\affiliation{Harish-Chandra Research Institute, A CI of Homi Bhabha National Institute, Chhatnag Road, Jhusi,
Prayagraj (Allahabad) 211019, India}

\author{Ujjwal Sen}
\email{ujjwal@hri.res.in}
\affiliation{Harish-Chandra Research Institute, A CI of Homi Bhabha National Institute, Chhatnag Road, Jhusi,
Prayagraj (Allahabad) 211019, India}

% Date
%\date{\today}

% Abstract
\begin{abstract}
We demonstrate the existence of a dynamical quantum phase transition (DQPT) in a dissipative collective-spin model that exhibits the boundary time crystal (BTC) phase. We initialize the system in the ground state of the Hamiltonian in either the BTC or the non-BTC phase, and drive it across the BTC transition. The driving is done by an abrupt quench or by a finite-time linear ramp of a Hamiltonian control parameter under Markovian Lindblad dynamics. We diagnose DQPTs through zeros of the fidelity-based Loschmidt echo between the initial state and the evolving mixed state, which induce nonanalytic cusp-like features in the associated rate function. For quenches into the BTC phase, the Loschmidt echo exhibits repeated zeros due to the emergent time-periodic steady state, whereas for quenches into the non-BTC phase, the overlap vanishes and remains zero once the dynamics relaxes to a stationary state. We further show that the DQPT persists under the ramp protocol followed by unitary evolution with the final Hamiltonian. Finally, we analyze the finite-size scaling of the first critical time and find convergence to a constant in the thermodynamic limit, with distinct power-law approaches for the quench and the ramp protocols.
\end{abstract}

% Keywords

% Make title
\maketitle

% Introduction
\section{Introduction}

Time crystals are non-equilibrium phases of matter in which time-translation symmetry is spontaneously broken. The original proposal of a time crystal in thermal equilibrium by  Wilczek~\cite{PhysRevLett.109.160401} was ruled out by no-go theorems~\cite{PhysRevLett.110.118901, PhysRevLett.111.070402, PhysRevLett.114.251603}, establishing that spontaneous breaking of continuous time-translation symmetry cannot occur in thermal equilibrium.
Nevertheless, the time-crystal phase can be realized away from equilibrium in both periodically driven closed and open many-body quantum systems~\cite{Sacha_2018, Khemani:2019nzi, Else_2020, RevModPhys.95.031001, PhysRevLett.117.090402}. {Time crystals are broadly classified into two categories, viz., discrete time crystals (DTC) and continuous time crystals (CTC)
%\textcolor{magenta}{[or boundary time crystals (BTC)]} 
based on whether a discrete or a continuous time translation symmetry is spontaneously broken in the time crystal phase.} In the widely studied setting of DTC, a periodically driven (Floquet) system spontaneously breaks the discrete time-translation symmetry of the drive Hamiltonian, typically via mechanisms such as many-body localization~\cite{PhysRevLett.117.090402, Zhang:2016kpq, PhysRevLett.116.250401, PhysRevB.94.085112, Ho:2017nee,Randall:2021ggc}  or prethermalization~\cite{PhysRevX.7.011026, Kuwahara:2016lfr, Abanin_2017, PhysRevB.95.014112, PhysRevX.10.021046}. 

{In a conceptually distinct route to time-crystalline order, the so-called boundary time crystals (BTCs) arise in {time-independent} open quantum many-body systems {with time-independent dynamical generator} in which persistent oscillations appear in the long-time steady-state dynamics without any external periodic drive. This phenomenon was first identified in Ref.~\cite{PhysRevLett.121.035301}, where the interplay of Hamiltonian interactions and {collective} dissipation generates a time-periodic steady state {despite the fact that the Hamiltonian is time independent}. } Necessary conditions responsible for the emergence of BTCs have been identified in Refs.~\cite{PhysRevB.104.014307, PhysRevB.106.224308}, and {a growing body of work has explored various aspects of BTCs and their applications}~\cite{PhysRevB.103.184308, Carollo:2021tiq, Lourenco:2021qbl, Rubio-Garcia:2022ebo, Cabot:2022bqp, Montenegro:2023jth, Carollo:2023nhz, Cabot:2023rqe, Gribben:2024uwx, Paulino:2024qcq, Passarelli:2025igu, Wang:2025nwc, Cabot:2025gdc, Nemeth:2025xgp, Das:2025pbp, OConnor:2025zrt, Liu:2025icd, Jirasek:2025cbe}. In particular, BTCs have been shown to provide advantages for quantum metrology and sensing~\cite{Montenegro:2023jth, Cabot:2023rqe, Gribben:2024uwx, Cabot:2025gdc, OConnor:2025zrt, Jirasek:2025cbe} and have also been investigated from a quantum-thermodynamic perspective~\cite{Carollo:2023nhz, Paulino:2024qcq}.

{In parallel, {another} major development in nonequilibrium quantum many-body physics is the notion of a dynamical quantum phase transition (DQPT) \cite{PhysRevLett.110.135704,Heyl:2017blm}. {Quantum phase transitions~\cite{Sachdev:2011fcc, RevModPhys.69.315} are a central paradigm in the theory of quantum many-body systems at equilibrium. They occur at zero temperature and are driven by quantum fluctuations, manifesting as nonanalytic behavior of ground-state properties when a Hamiltonian control parameter is tuned across a critical value in the thermodynamic limit. While equilibrium quantum phase transitions are by now well understood, much less is known about their dynamical counterparts that arise during nonequilibrium time evolution. These DQPTs are characterized by nonanalytic structures emerging in real time during unitary or dissipative evolution following a quench or ramp of system parameters.} In the standard closed-system quench protocol, the system is initialized in the ground state of an initial Hamiltonian, following which the Hamiltonian is subjected to a sudden change in one of its parameters. Subsequently, the system evolves with the time-independent post-quench Hamiltonian. The return probability to the initial state, quantified by the Loschmidt echo, can develop zeros at critical times, signaling the onset of DQPT. Correspondingly, the associated rate function (often interpreted as a dynamical free-energy density) develops cusp-like nonanalyticities at those critical times in the thermodynamic limit. DQPTs have been observed experimentally in a variety of platforms~\cite{Flaschner:2016gju, Zhang:2017kde, PhysRevLett.119.080501, PhysRevB.100.024310, PhysRevApplied.11.044080, PhysRevLett.122.020501, PhysRevA.100.013622, PhysRevLett.124.043001, PhysRevLett.124.250601, Dborin:2022zdd, Karch:2025wli}. For reviews on DQPT, see Refs.~\cite{Heyl:2017blm, zvyagin2017dynamicalquantumphasetransitions, Heyl:2018jzi}. Importantly for the present work, DQPTs also occur in open quantum systems, where the Loschmidt echo can be generalized using fidelity between mixed states~\cite{Bandyopadhyay:2018zgm}.}

{These developments motivate a natural question: how do DQPTs manifest in intrinsically non-equilibrium time-crystalline phases that are stabilized by dissipation? While DQPTs have been explored in a broad range of dissipative models~\cite{Bandyopadhyay:2018zgm, PhysRevB.98.134310, PhysRevA.101.012111, PhysRevLett.100.040403, PhysRevLett.125.143602, PhysRevA.105.022220, DiBello:2023qkw, PhysRevB.108.075110, PhysRevB.97.045147, Zhang:2025eze, Parez:2025psa}, their interplay with the BTC phase, where the late-time state is time-periodic rather than stationary, has not been explored. In this context, it has been shown that discrete time crystals in closed quantum systems can host DQPTs~\cite{PhysRevA.97.053621}. In this work, we demonstrate the existence of DQPTs in a dissipative collective-spin model that exhibits the BTC phase under Markovian Lindblad dynamics. We initialize the system in the ground state of the Hamiltonian in one phase (BTC or non-BTC), and then drive the system across the BTC transition using either (i) an abrupt quench or (ii) a finite-time linear ramp of a Hamiltonian control parameter. We diagnose DQPTs using zeros of the fidelity-based Loschmidt echo between the initial state and the time-evolved mixed state, which induce cusp-like nonanalyticities in the corresponding rate function.}

{Our results reveal a qualitative distinction tied to the nature of the late-time dynamics. For quenches into the BTC phase, the time-periodic steady state leads to repeated zeros of the fidelity-based Loschmidt echo at a sequence of critical times. In contrast, for quenches into the non-BTC phase, the dynamics relaxes to a stationary steady state, and the overlap with the pre-quench initial state can vanish and remain zero beyond the first critical time. We further show that the DQPT persists under ramp protocols followed by unitary evolution with the final Hamiltonian. Finally, in both the quench and ramp protocols, we analyze finite-size scaling of the first critical time and find convergence to a constant in the thermodynamic limit, with distinct power-law approaches for quench and ramp protocols.}

{The remainder of the paper is organized as follows. In Sec.~\ref{sec:prelim} we briefly review BTCs and DQPTs, including the fidelity-based Loschmidt echo appropriate for diagnosing DQPTs in open-system dynamics. In Sec.~\ref{sec:setup}, we describe the collective-spin model and the quench and the ramp protocols used to probe DQPTs across the BTC transition. In Sec.~\ref{sec:results} we present our numerical results and finite-size scaling analysis. We conclude in Sec.~\ref{sec:conclusion}.}

\begin{figure*}[htp]
    \centering
    \begin{minipage}[t]{0.4\textwidth}
        \centering
        \includegraphics[trim=0cm 0.0cm 0.0cm 0, clip,width=7.5cm, height=4cm]{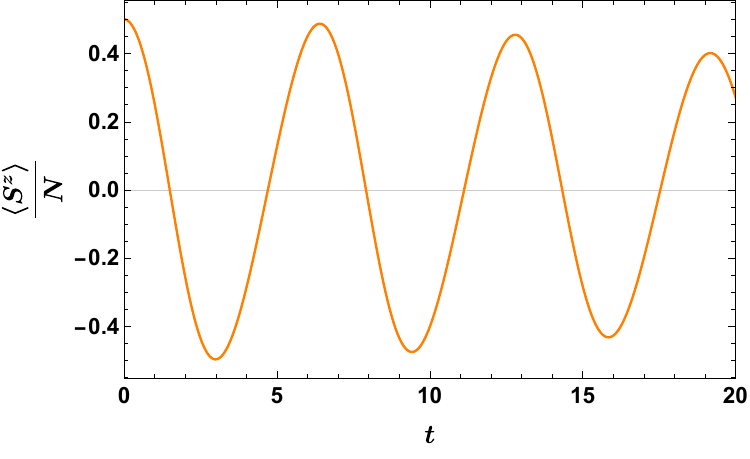}\\
        \textbf{(a)}
    \end{minipage}
    \hspace{1 cm}
    \begin{minipage}[t]{0.4\textwidth}
        \centering
        \includegraphics[trim=0cm 0.0cm 0.0cm 0, clip, width=7.5cm, height=4cm]{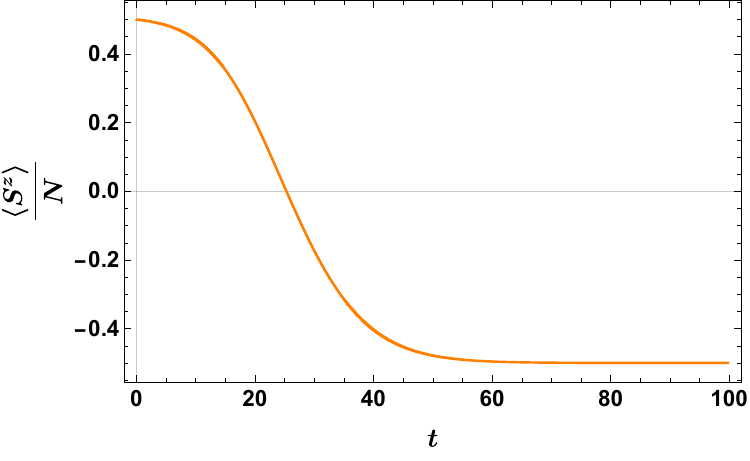}\\
        \textbf{(b)}
    \end{minipage}   \caption{\textit{Illustration of BTC and non-BTC phases in the model of Eq.~\eqref{eq:master-eqn}.} Shown is the average magnetization $\langle S^z \rangle/N$ for $N=100$ as a function of time in (a) the BTC phase and (b) the non-BTC phase. In the BTC regime, the dynamics of the order parameter $\langle S^z \rangle/N$ relaxes to a time-periodic steady state with persistent oscillations, whereas in the non-BTC regime, the average magnetization approaches a time-independent stationary value. The parameters corresponding to the BTC phase are chosen as $\omega_0 = -1$, $\omega_x = 0$, $\omega_z = -0.25$, and $\kappa = 0.1$. For the non-BTC phase, we set $\omega_0 = 0$, while keeping all other parameters unchanged. The quantities plotted along the horizontal and vertical axes in each panel are dimensionless. }
    \label{fig:Sz}
\end{figure*}

\section{Preliminaries}\label{sec:prelim}

\subsection{Boundary Time Crystals}

{The boundary time crystal (BTC) is an intrinsically non-equilibrium phase of open many-body systems in which the long-time dynamics exhibits persistent oscillations despite a time-independent generator of evolution. Concretely, in the BTC phase, the system relaxes to a time-periodic steady state, so that expectation values of suitable observables display stable oscillations at late times. In a Markovian description, this behavior is naturally characterized via the spectrum of the Liouvillian superoperator. The emergence of undamped oscillations is associated with purely imaginary Liouvillian eigenvalues, leading to time-translation-symmetry breaking in the steady-state dynamics.} The BTC phase was first identified in Ref.~\cite{PhysRevLett.121.035301} in the context of collective spin models originally introduced to describe cooperative emission of radiation in cavity-QED type settings~\cite{PhysRevA.98.042113, 10.1143/PTPS.64.307, DRUMMOND1978160, PURI1979200, DFWalls_1980, PhysRevA.65.042107}.

The BTC phase can be diagnosed operationally by (i) persistent oscillations of collective observables (e.g. average magnetization) at late times, and (ii) the corresponding Liouvillian spectral signature mentioned above. {As a representative example}, we consider the collective-spin model of Ref.~\cite{PhysRevB.103.184308} comprising $N$ spin-$1/2$ particles, whose Hamiltonian and Lindblad operator take the form,
\begin{align}
H &= K \left( \omega_0 S^x + \frac{\omega_x}{S} \left( S^x\right)^2 + \frac{\omega_z}{S} \left( S^z\right)^2 \right),  \label{BTC-Ham}\\
L &= \sqrt{\frac{K \kappa}{S}} S_{-} \label{BTC-Lin}
\end{align}
where $S = N/2$ is the total spin of the system, $S^\alpha = \sum_{j=1}^{N} \sigma^\alpha_j/2$ with $\alpha =x,y,z$ are collective spin operators, $S_\pm = S^x \pm \mathrm{i} S^y$ and $\sigma^\alpha_j$ are the Pauli spin operators for the $j$th spin.
%~\textcolor{magenta}{[CITE]}. \sout{In the presence of the interaction terms, the phase diagram is more involved as described in Ref.}~\cite{PhysRevB.103.184308}~\textcolor{magenta}{[are you sure that you want to cite Ref.~\cite{PhysRevB.103.184308}  here?]}, \sout{which nevertheless supports the BTC phase over wide parameter ranges.} 
{Note that $K$ is a constant parameter with units of energy. In this paper, we express all the physical quantities in units of $K=1$. All the parameters $\omega_0$, $\omega_x$, $\omega_z$, and $\kappa$ are dimensionless.}
{For appropriate parameter regimes, this model exhibits a BTC phase in which the long-time dynamics displays robust oscillations of collective observables, whereas outside that regime the system relaxes to a stationary time-independent steady state~\cite{PhysRevB.103.184308}. In what follows, we refer to these regimes as the BTC and non-BTC phases, respectively.}

\subsection{Dynamical Quantum Phase Transitions}\label{sec:prelim::subsec:DQPT}

Dynamical quantum phase transitions (DQPTs)~\cite{PhysRevLett.110.135704, Heyl:2017blm} are characterized by the non-analyticities in the real-time evolution of a quantum many-body system, induced by a time-independent final Hamiltonian after either an abrupt quench or a gradual change in a Hamiltonian parameter. 
Consider a quantum many-body system of size $N$ {described by a Hamiltonian $H$}, initially prepared in the ground state {of $H$}, $| \psi (0) \rangle$  at time $t=0$. {Note that $t$ denotes dimensionless time which is related to the actual time $\tilde{t}$ by $t = \frac{K \tilde{t}}{\hbar}$.} At this instant, a control parameter $\lambda$ in the Hamiltonian is abruptly quenched from an initial value $\lambda_i$ to a final value $\lambda_f$. The subsequent dynamics is governed by the time-independent Hamiltonian $H(\lambda_f)$, such that the state at time $t$ evolves as $| \psi(t) \rangle = \exp \left( - \mathrm{i} H(\lambda_f) t\right) | \psi (0) \rangle$. A DQPT is said to occur at critical times for which the evolved state becomes orthogonal to the initial state. Equivalently, these critical times are marked by the vanishing of the Loschmidt echo (LE), $\mathcal{L}:= |\langle \psi (0)|\psi(t)  \rangle|^2$. At those times, the associated dynamical free-energy density,
or the rate function of the return probability, $f := - \lim_{N \to \infty} (1/N) \ln \mathcal{L}$, diverges. Such non-analyticities also appear when the Hamiltonian control parameter is changed via a slow ramping protocol instead of sudden quenching~\cite{10.21468/SciPostPhys.1.1.003, PhysRevB.93.144306}. 

DQPTs have also been shown to exist under dissipative dynamics~\cite{Bandyopadhyay:2018zgm, PhysRevB.98.134310, PhysRevA.101.012111, PhysRevLett.100.040403, PhysRevLett.125.143602, PhysRevA.105.022220, DiBello:2023qkw, PhysRevB.108.075110, PhysRevB.97.045147, Zhang:2025eze, Parez:2025psa}.
{For open-system dynamics, the Loschmidt echo is naturally generalized using the Uhlmann fidelity~\cite{UHLMANN1976273, Jozsa:1994qja}. For two density matrices $\rho$ and $\sigma$, the Uhlmann fidelity is}
\begin{equation}\label{eq:Uhlmann-Fid}
F(\rho,\sigma) := \left[\mathrm{Tr}\sqrt{\sqrt{\rho}\,\sigma\,\sqrt{\rho}}\right]^2 .
\end{equation}
We define the fidelity-based Loschmidt echo (FLE) as
\begin{equation}\label{eq:FLE}
    \mathcal{L}_F(t) := F\!\left(\rho(0),\rho(t)\right),
\end{equation}
and the corresponding rate function as
\begin{equation}
f_F(t) := - \lim_{N \to \infty} \frac{1}{N}\ln \mathcal{L}_F(t).
\label{eq:rateF_def}
\end{equation}
DQPT critical times are then identified by zeros of $\mathcal{L}_F(t)$, at which $f_F(t)$ develops cusp-like singularities. In the special case where the initial state is pure, $\rho(0)=|\psi (0)\rangle\langle\psi (0)|$, Eq.~\eqref{eq:FLE} reduces to 
\begin{equation}
\mathcal{L}_F(t)=\langle\psi(0)|\rho(t)|\psi(0)\rangle .
\label{eq:FLE_pure_simplify}
\end{equation}
In this work, we use $\mathcal{L}_F(t)$ and $f_F(t)$ as diagnostics of DQPTs under Lindblad evolution, and we consider both abrupt quenches and finite-time ramp protocols across the BTC transition.

\section{Setup}\label{sec:setup}

\begin{figure*}[t]
    \centering
    \begin{minipage}[t]{0.3\textwidth}
        \centering
        \includegraphics[trim=0cm 0.0cm 0.0cm 0, clip,width=5.5cm, height=4cm]{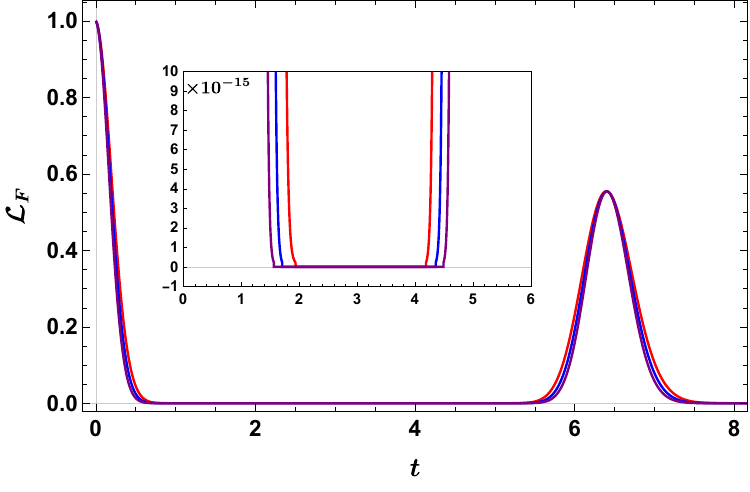}\\
        \textbf{(a)}
    \end{minipage}
    \hspace{0.5 cm}
    \begin{minipage}[t]{0.3\textwidth}
        \centering
        \includegraphics[trim=0cm 0.0cm 0.0cm 0, clip, width=5.5cm, height=4cm]{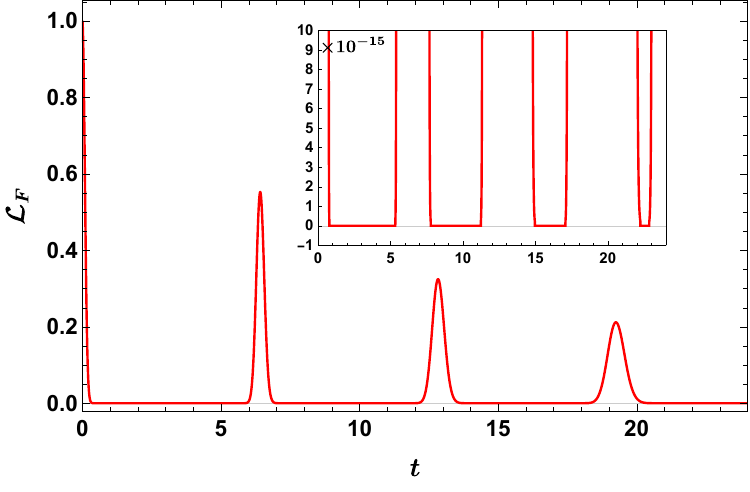}\\
        \textbf{(b)}
    \end{minipage} 
    \hspace{0.5 cm}
    \begin{minipage}[t]{0.3\textwidth}
        \centering
        \includegraphics[trim=0cm 0.0cm 0.0cm 0, clip, width=5.5cm, height=4cm]{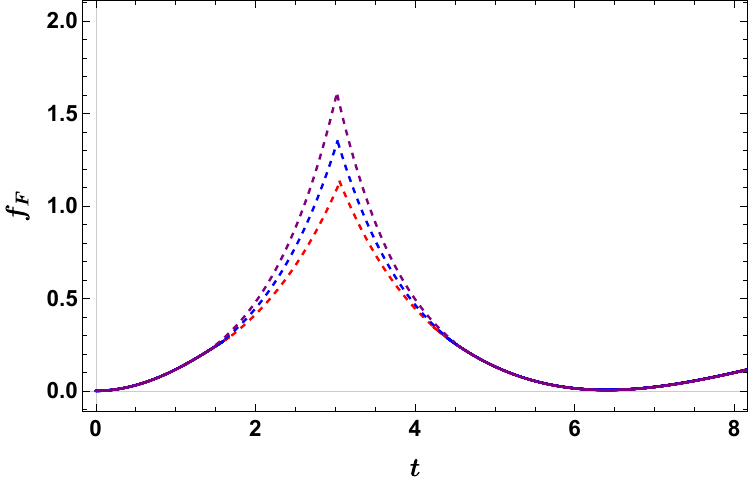}\\
        \textbf{(c)}
    \end{minipage}\caption{\textit{Dynamical quantum phase transition under a quench from the non-BTC to the BTC phase.} \textbf{(a)} The fidelity-based Loschmidt echo $\mathcal{L}_F(t)$ is shown as a function of time following an abrupt quench of $\omega_0$ from $\omega_{0,i} = 0$ (non-BTC) to $\omega_{0,f} = -1$ (BTC) for system sizes $N=50$ (red), $N=60$ (blue), and $N=70$ (purple). The system is initially prepared in the ground state of the pre-quench Hamiltonian and subsequently evolves under the Lindblad dynamics with the post-quench Hamiltonian. The first vanishing of $\mathcal{L}_F(t)$ defines the first critical time $\tco$, signalling the onset of a DQPT. The inset magnifies the region near the $t$-axis to highlight the numerical zeros. The numbers smaller than machine precision ($10^{-15}$) are concurrent with zero. We obtain $\tco$ as the midpoint of the first times at which $\mathcal{L}_F$ drops below and rises above $10^{-15}$, denoted respectively as $\tmo$ and $\tpo$. The values of the first critical times, $\tco$, along with the respective values of $\tmo$ and $\tpo$ for each of the three $N$ are provided in Tab.~\ref{tab:FLE-quench}. 
    The data is generated with the time-step of $\Delta t = 0.001$. \textbf{(b)} Plot of $\mathcal{L}_F(t)$ for $N=250$ with same parameters as in (a) demonstrating that $\mathcal{L}_F(t)$ periodically becomes zero. \textbf{(c)} The rate function $f_F(t) = -(1/N) \ln \mathcal{L}_F(t)$ corresponding to (a) is shown with the curves of the same colors for the respective $N$ values as in (a).
    %for $N=50$ (red), $N=60$ (blue), $N=70$ (purple).
    At the first critical time $\tco$, where $\mathcal{L}_F$ vanishes, $f_F$ exhibits a cusp-like singularity, signalling a DQPT. Solid curves show numerically accessible values of $f_F$, while dashed curves are extrapolations inside the region where $\mathcal{L}_F$ falls below machine precision. }
    \label{fig:FLE-quench}
\end{figure*}

{We consider a system of $N$ spin-1/2 particles described by a {collective-spin} Hamiltonian $H$ of Eq.~\eqref{BTC-Ham} {which evolves under Markovian dynamics with} a single jump operator $L$ given in Eq.~\eqref{BTC-Lin}. The Markovian dynamics is governed by the Lindblad master equation,}
%We consider a system of $N$ spin-1/2 particles whose Hamiltonian and Lindblad operator are given by Eq.~\eqref{BTC-Ham} and~\eqref{BTC-Lin}, respectively, and whose dynamics is governed by the Lindblad master equation,
\begin{align}\label{eq:master-eqn}
    \frac{d \rho }{d t}=\mathscr{L}[\rho] &= -\mathrm{i} [H, \rho] +  \mathcal{D}[\rho], \nonumber \\
    \qquad \mathcal{D}[\rho] &= L \rho L^\dagger - \frac{1}{2} \{L^\dagger L, \rho \}.
\end{align}
This model supports both a BTC regime, in which the long-time steady state is time periodic, and a non-BTC regime, in which the dynamics relaxes to a stationary steady state~\cite{PhysRevB.103.184308}.
%This model is shown to exhbit BTC phase over a wide range of parameters~\cite{PhysRevB.103.184308}. 
{In the BTC phase, the average magnetization, $\langle S^z \rangle/N $, varies periodically in time, and the spectrum of the Liouvillian $\mathscr{L}$ contains purely imaginary eigenvalues. In the BTC phase, such time-periodic behavior of $\langle S^z \rangle/N $ exists for all the initial states and is robust against perturbations in Hamiltonian parameters and noise~\cite{PhysRevB.103.184308}.}
%\SM{Shall we give a plot of $\langle S^z \rangle/N $ as a function of time?}

To probe DQPTs across the BTC transition, we consider two standard driving protocols.
%In order to demonstrate the existence of DQPT in this model we follow two standard protocols. 
In both, the system is initialized in the ground state $\rho(0) = | \psi(0) \rangle \langle \psi(0) |  $ of the Hamiltonian $H(\lambda_i)$. We then vary a single Hamiltonian parameter $\lambda \in \{ \omega_0, \omega_x, \omega_z \}$ either abruptly (quench protocol) or slowly (ramp protocol) from an initial value $\lambda_i$ to a final value $\lambda_f$, while keeping the remaining parameters fixed.
%In both the protocols, initial state is prepared as the ground state of Hamiltonian~\eqref{BTC-Ham} for a fixed choice of Hamiltonian parameters following which one of the parameters of the Hamiltonian is changed either abruptly (quench protocol) or slowly (ramp protocol). 
This change in Hamiltonian parameter results in the time evolution of the initial state according to the master equation~\eqref{eq:master-eqn} with the final Hamiltonian $H(\lambda_f)$. A DQPT is diagnosed by zeros of the fidelity-based Loschmidt echo $\mathcal{L}_F(t)$ introduced in Sec.~\ref{sec:prelim::subsec:DQPT}, and also equivalently by cusp-like nonanalyticities in the corresponding rate function $f_F(t)$.
%This time evolution under new Hamiltonian is characterized by the vanishing of the FLE at critical times signalling DQPT. 
{Note that the change in parameter $\lambda$ must be across the BTC phase transition point. In other words, if the initial Hamiltonian parameter choice corresponds to the non-BTC phase, then the final Hamiltonian parameters after quench/ramp should correspond to the BTC phase, and vice versa.}

In our 
%numerical results 
work (Sec.~\ref{sec:results}), we take $\lambda = \omega_0$ for concreteness; {however, the protocols described
%, we apply in our work, is discussed below.
below apply to varying any of the Hamiltonian parameters.}

\subsection{Quench Protocol}

In the quench protocol~\cite{PhysRevLett.110.135704, Bandyopadhyay:2018zgm}, after preparing the initial ground state $| \psi(0) \rangle$ of $H(\lambda_i)$, 
%as the ground state of the Hamiltonian~\eqref{BTC-Ham} with parameter $\lambda_{i}$ at initial time $t=t_i$, $\lambda$ is suddenly changed to $\lambda_f$. 
we abruptly change $\lambda_i \to \lambda_f$ at $t=0$.
%Subsequently $| \psi(0) \rangle$ evolves according to the master equation~\eqref{eq:master-eqn} with Hamiltonian parameter $\lambda_f$. 
The state then evolves under Eq.~\eqref{eq:master-eqn} with the time-independent post-quench Hamiltonian $H(\lambda_f)$. We compute $\mathcal{L}_F(t) = F(\rho(0), \rho(t))$ {whose first zero defines the first critical time and} signals the onset of a DQPT.

\subsection{Ramp Protocol}
\label{sec:setup::subsec:ramp}
{We also consider a finite-time ramp protocol introduced in Ref.~\cite{Bandyopadhyay:2018zgm} for detecting DQPT in open quantum systems.} In the ramp protocol~\cite{PhysRevB.93.144306, Bandyopadhyay:2018zgm}, instead of abrupt quenching, the parameter $\lambda$ is varied smoothly from $\lambda_i$ to $\lambda_f$ over a duration $\tau$ as 
\begin{equation}\label{eq:ramp}
    \lambda(t) = \lambda_i + (\lambda_f - \lambda_i) \frac{t}{\tau}, \qquad 0 \leq t \leq \tau
\end{equation}
%
%where $\tau$ is the duration of the ramp. For the time $t_i < t < \tau$, 
During the ramp ($0 \leq t \leq \tau$), the system evolves according to Eq.~\eqref{eq:master-eqn} with a time-dependent Hamiltonian $H(\lambda(t))$.
%, where parameter $\lambda$ changes in time according to Eq.~\eqref{eq:ramp}. 
At the end of the ramp, $t = \tau$, the dissipation is switched off, and subsequently the state evolves unitarily with the final Hamiltonian $H(\lambda_f)$. 
Treating $\rho(\tau)$ as the initial state for this post-ramp unitary stage, we compute 
\begin{equation}\label{eq:LF-ramp}
    \mathcal{L}_F(t') = F(\rho(\tau), \rho(\tau + t')), \qquad t' \geq 0
\end{equation}
and identify DQPT critical times by zeros of $\mathcal{L}_F(t')$ (equivalently cusps in $f_F(t')$).

\section{DQPT Across the BTC Transition}\label{sec:results}
In this section, we demonstrate the emergence of DQPTs in the dissipative collective-spin model of Eq.~\eqref{eq:master-eqn} when a Hamiltonian control parameter is driven across the BTC/non-BTC phase boundary. We diagnose DQPTs using the fidelity-based Loschmidt echo $\mathcal{L}_F(t)$ and its rate function $f_F(t)$ introduced in Sec.~\ref{sec:prelim::subsec:DQPT}. Critical times correspond to zeros of $\mathcal{L}_F(t)$, at which $f_F(t)$ develops nonanalytic cusp-like singularities in the thermodynamic limit.

Throughout this section we vary $\lambda = \omega_0$ while keeping $\omega_x = 0$, $\omega_z = -0.25$, and $\kappa = 0.1$ fixed. For the non-BTC $\to$ BTC protocols we choose $\omega_{0,i} = 0$ (non-BTC) and $\omega_{0,f} = -1$ (BTC), while for the BTC $\to$ non-BTC ones we choose $\omega_{0,i} = -1$ (BTC) and $\omega_{0,f} = 0$ (non-BTC). The corresponding long-time behavior of the order-parameter observable, namely, average magnetization  $\langle S^z \rangle/ N$ is shown in Fig.~\ref{fig:Sz} for $N=100$. In the BTC regime, the magnetization exhibits persistent oscillations in time, whereas in the non-BTC regime, it relaxes to a time-independent stationary value.

\paragraph*{\textit{Numerical convention for identifying zeros.}} In finite-precision numerics, $\mathcal{L}_F(t)$ does not typically hit exact zero. We therefore treat values below a threshold $\varepsilon$ as numerically zero, and define the first critical time $\tco$ as the midpoint
\begin{equation}
    \tco = \frac{\tpo + \tmo}{2}
\end{equation}
where $\tpo$ ($\tmo$) is the first time at which $\mathcal{L}_F(t)$ drops below (rises above) $\varepsilon$. For quenches, we can take $\varepsilon = 10^{-15}$ (machine precision), while for the ramp protocol, we use $\varepsilon = 10^{-13}$ due to enhanced numerical noise in mixed-state fidelity evaluations for large $N$. We use a time step $\Delta t = 0.001$  throughout.

\subsection{Quench Protocol}
Let us first discuss the quench protocol from the stationary non-BTC regime to the BTC regime. Fig.~\ref{fig:FLE-quench}~(a) displays a plot of $\mathcal{L}_F(t) =F(\rho(0), \rho(t))$ as a function of time for $N=50$ (red), $N=60$ (blue), and $N=70$ (purple). The first interval where $\mathcal{L}_F(t)$ falls below $10^{-15}$ defines $\tmo$ and $\tpo$, and the resulting $\tco$ values are reported in Tab.~\ref{tab:FLE-quench}.

\begin{table}[t]
\centering
\setlength{\tabcolsep}{12pt}   % ← increase from default (~6pt)
%\begin{tabular*}{\columnwidth}{|c|c|c|c|@{\extracolsep{\fill}}}
\begin{tabular}{|c|c|c|c|}
\hline
\textbf{$N$} & \textbf{$\tmo$} & \textbf{$\tpo$} & \textbf{$\tco$} \\
\hline
 50 & 1.869 & 4.232 & 3.051 \\
\hline
 60 & 1.666 & 4.398 & 3.032 \\
\hline
 70 & 1.522 & 4.530 & 3.026 \\
\hline
\end{tabular}
%\end{tabular*}
\caption{\textit{Quench protocol: non-BTC $\to$ BTC}. The values of $\tmo$, $\tpo$, and $\tco$ in Fig.~\ref{fig:FLE-quench}~(a) corresponding to $N=50$, $60$, and $70$.}
\label{tab:FLE-quench}
\end{table}

Since the initial state is the pure ground state $\rho(0) = | \psi(0) \rangle \langle  \psi(0) |$ of $H(\omega_{0,i})$, the fidelity-based Loschmidt echo simplifies to the overlap
\begin{equation}
    \mathcal{L}_F(t) = \langle \psi(0) | \rho(t) | \psi(0) \rangle
\end{equation}
The corresponding rate function $f_F(t)$ is plotted in {Fig.~\ref{fig:FLE-quench}~(c)}. At the first critical time $\tco$, the echo vanishes (within numerical precision), and $f_F(t)$ diverges accordingly. Within the accessible numerical precision, we are able to numerically compute $f_F$ for the values of $t$ for which $\mathcal{L}^F > 10^{-15}$ (machine precision). Since for the values of $t$ for which $\mathcal{L}_F  \leq 10^{-15}$, we set $\mathcal{L}_F$ to zero, the value of $f^F$ for all these times is divergent within the numerical precision. The solid curves in {Fig.~\ref{fig:FLE-quench}~(c)} depict $f_F$ for the times where $\mathcal{L}_F > 10^{-15}$. We extrapolate this $f_F$ to the region where $\mathcal{L}_F \leq 10^{-15}$ and represent these extrapolated curves by dashed lines of respective colors. This extrapolation is done by fitting the curve to the left of $\tco$ by an exponential function of the form, $a_L e^{b_L (t - \tco) }+ c_L$ where $a_L$, $b_L$, and $c_L$ are fitting parameters, and fitting the curve to the right of $\tco$ by another exponential function of the form $a_R e^{b_R ( \tco - t)} + c_R$, where $a_R$, $b_R$, and $c_R$ being the corresponding fitting parameters. The two extrapolated curves meet at $t = \tco$, culminating in a visible cusp. Note that this extrapolation to display the cusp is solely for the representative purpose.

A qualitative feature specific to quenches into the BTC phase is that zeros of $\mathcal{L}_F(t)$ also occur at times later than $\tco$. Because the long-time dynamics approach a time-periodic steady state in the BTC regime, the evolving state can repeatedly become orthogonal to the initial state at a sequence of critical times, while returning to nonzero overlap at intermediate times. {This behavior is visible in Fig.}~\ref{fig:FLE-quench}~(b) which depicts the plot of $\mathcal{L}_F(t)$ for $N=250$ with the same parameter values as in Fig.~\ref{fig:FLE-quench}~(a).

\begin{figure}[htbp]
    \centering
    \includegraphics[width=1\linewidth]{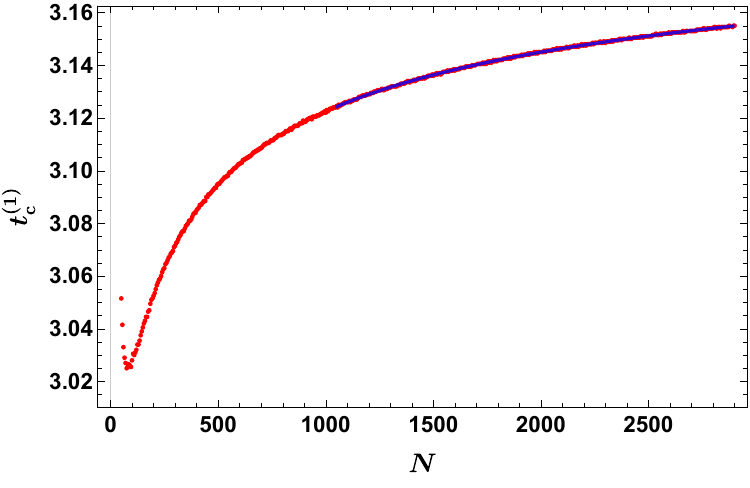}
    \caption{
    %\textbf{Quench protocol: large-$N$ scaling of $t_c$} (non-BTC to BTC)- 
    \textit{Finite-size scaling of the first critical time under a quench from the non-BTC to the BTC phase.}
    The first critical time $\tco$ is plotted as a function of system size $N$.
    The red dots are the numerical data, while the blue curve is a power-law fit of the form $\tco(N) =  a N^{-b} + c$ in the range $N \in [1045,2900]$, indicating convergence to a constant value in the thermodynamic limit. 
    %Fit parameters are given in the text.
    The best-fit values of fitting parameters are $a = -1.194 \pm 0.053 $, $b=0.354 \pm 0.009$, and $c=3.225 \pm 0.002 $. Since the data is generated with a time resolution of $\Delta t = 0.001$, values of the parameters up to the third decimal place are meaningful. Therefore, we have truncated the errors to the third decimal place. The RMS fitting error is less than 0.001.
    The values of both pre- and post-quench parameters are the same as in Fig.~\ref{fig:FLE-quench}.
    %First critical time, $\tco$, is plotted as a function of system size $N$ for the quench protocol. 
    %$\omega_0$ is quenched from $\omega_{0,i} = 0$ to $\omega_{0,f} = -1$. In this case, the quenching is from the non-BTC phase to the BTC phase. The initial state is the ground state of Hamiltonian~\eqref{BTC-Ham} with parameter values $\omega_0 = \omega_{0,i} = 0$, $\omega_x =0$, $\omega_z = -0.25$. Following the quench the initial state evolves according to the master equation~\eqref{eq:master-eqn} with the final Hamiltonian having $\omega_0 = \omega_{0,f} = -1$ and all other parameters unchanged and Lindblad operator~\eqref{BTC-Lin} with $\kappa = 0.1$. 
    %The values of both pre- and post-quench parameters are the same as in Fig.~\ref{fig:FLE-quench}.
    %The first critical time is the time at which $\mathcal{L}^F$ first becomes zero signalling the onset of DQPT. 
    %The solid red curve depicts the behavior of $\tco$ with the system size $N$. The dashed blue curve is a power-law fit of the form $t_c(N) =  a N^{b} + c$ in the range $N \in [1045,2900]$ of the $t_c(N)$ data used to generate the solid red curve. The best-fit values of parameters are $a = -1.194 \pm 0.053 $, $b=-0.354 \pm 0.009$, and $c=3.225 \pm 0.002 $. Since the data is generated with the time resolution of $\Delta t = 0.001$, values of the parameters upto third decimal place are meaningful. Therefore we have truncated the errors to third decimal place. The RMS fitting error is zero upto the third decimal place.
    }
    \label{fig:tc-scaling-N-quench}
\end{figure}

\begin{figure}[htbp]
    \centering
    \includegraphics[width=1\linewidth]{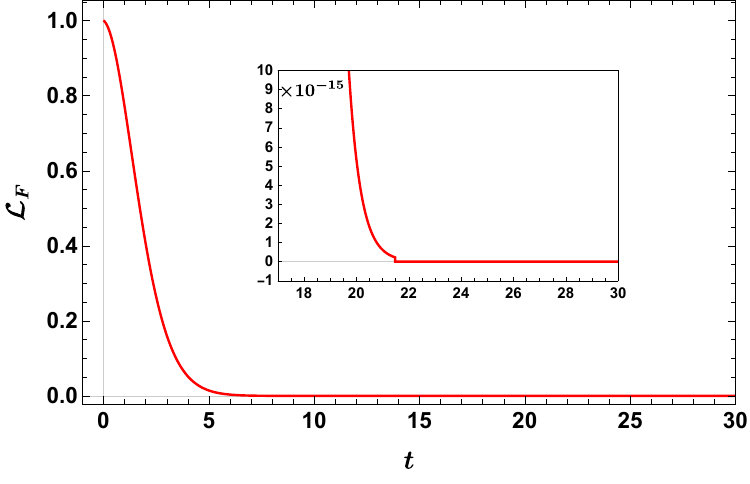}
    \caption{
    \textit{DQPT under a quench from the BTC to the non-BTC phase.}
    %\textbf{Quench protocol: zeros of $\mathcal{L}^F$} (BTC to non-BTC)- 
    The fidelity-based Loschmidt echo $\mathcal{L}_F(t)$ is shown for $N=100$ following a quench of $\omega_0$ from $\omega_{0,i} = -1$ (BTC) to $\omega_{0,f} = 0$ (non-BTC). After the first vanishing of $\mathcal{L}_F$ at $\tco = 20.778$, the overlap remains zero as the dynamics relaxes to a stationary steady state. The inset highlights the behavior near the $t$-axis. Unlike quenches into the BTC phase, no revivals of the overlap occur at later times.
    %$\mathcal{L}^F (\rho(0), \rho(t))$ is plotted as a function of time following a quench protocol for $N=100$. In this case, the quenching is from the BTC phase to the non-BTC phase. In particular, $\omega_0$ is quenched from $\omega_{0,i} = -1$ (BTC) to $\omega_{0,f} = 0$ (non-BTC). The initial state is the ground state of Hamiltonian~\eqref{BTC-Ham} with parameter values $\omega_0 = \omega_{0,i} = -1$, $\omega_x =0$, $\omega_z = -0.25$. Following the quench, the initial state evolves according to the master equation~\eqref{eq:master-eqn} with the final Hamiltonian having $\omega_0 = \omega_{0,f} = 0$ and all other parameters unchanged, and Lindblad operator~\eqref{BTC-Lin} with $\kappa = 0.1$. The time at which $\mathcal{L}^F$ drops below machine precision, $10^{-15}$, is $\tco = 20.778$, signalling the onset of DQPT. The inset shows the region zoomed in near the $t$-axis. The numbers smaller than the machine precision ($10^{-15}$) are concurrent with zero. The data is generated with the time-step of $\Delta t = 0.001$.
    }
    \label{fig:FLE-quench-BTC-to-nBTC}
\end{figure}

Furthermore, we have investigated the scaling of $\tco$ with the system size $N$ and found the asymptotic value of $\tco$ in the thermodynamic limit, $N \to \infty$. The plot of $\tco$ as a function of $N$ is shown in Fig.~\ref{fig:tc-scaling-N-quench} for the same values of parameters used in {Fig.~\ref{fig:FLE-quench}}. The red dots constitute the actual data generated for $N=50$ to $N=2900$ in steps of 5. We observe that $\tco$ initially decreases, followed by a subsequent increase, and it approaches a constant value asymptotically for large $N$. The blue curve is the best power-law fit of the form $\tco(N) = a N^{-b} + c$ in the range $ 1045 < N < 2900$ for the $\tco(N)$ data used to generate the red curve. The best-fit values of parameters are $a = -1.194 \pm 0.053 $, $b=0.354 \pm 0.009$, and $c=3.225 \pm 0.002 $. Since the data is generated with a time resolution of $\Delta t = 0.001$, values of the parameters up to the third decimal place are meaningful. Therefore, we have truncated the best-fit parameter values and their errors to the third decimal place. The root-mean-square (RMS) fitting error is zero up to the third decimal place. In other words, we find that in the $N \to \infty$ limit $\tco$ approaches a constant value of $3.225$ via a power-law with an exponent of $0.354$. 

We have also fitted the same data of Fig.~\ref{fig:tc-scaling-N-quench} with exponential and log functions. The corresponding values of fitted parameters and errors are reported in Tab.~\ref{tab:FLE-quench-tc1-scaling}. Observation that the exponential fit performs comparable to the power-law fit can be explained as follows. First note that $N^{-0.354}$ can be written as $e^{-0.354 \ln  N}$, whereas the exponential function goes as $e^{-N^{0.157}}$. Both $N^{0.157}$ and $0.354 \ln  N$  are close to each other for the range $1045 < N < 2900$ used for the fitting. This is why the performance of both the power-law and exponential fits is comparable. However, it can be easily seen that the function $N^{0.157}$ significantly dominates over $0.354 \ln  N$ for $N > 10^9$ and therefore in the thermodynamic limit $N \to \infty$, the two functions will have completely disparate asymptotic behaviors.

\begin{table}[t]
\centering
\begin{tabular}{|c|c|c|}
\hline
Fitting function & Fitted parameters & RMS error \\
\hline
\multirow{3}{*}{$a N^{-b} + c$}
 & $a = -1.194 \pm 0.053$ & \multirow{9}{*}{$< 10^{-3}$} \\ \cline{2-2}
 & $b = 0.354 \pm 0.009$ &  \\ \cline{2-2}
 & $c = 3.225 \pm 0.002$ &  \\
\cline{1-2}
\multirow{3}{*}{$a \exp(-N^b) + c$}
 & $a = -1.505 \pm 0.025$ &  \\ \cline{2-2}
 & $b = 0.157 \pm 0.001$ &  \\ \cline{2-2}
 & $c = 3.199 \pm 0.001$ &  \\
\cline{1-2}
\multirow{3}{*}{$a \ln(N) + c$}
 & $a = 0.029 \pm (<10^{-3})$ &  \\ \cline{2-2}
 & $c = 2.918 \pm (<10^{-3})$ &  \\ \cline{2-2}
\hline
\end{tabular}
\caption{\textit{Finite-size scaling of $\tco$ for quench protocol from non-BTC to BTC.} The fitting functions and corresponding best-fit values of fitting parameters, along with their errors, are shown. The last column shows the overall root-mean-square (RMS) fitting error. The data used for fitting is the data indicated by red dots in Fig.~\eqref{fig:tc-scaling-N-quench} within the range $1045 < N < 2900$. We have truncated all the numerical values to the third decimal place as the time resolution of our numerics is $\Delta t = 0.001$.}
\label{tab:FLE-quench-tc1-scaling}
\end{table}

Finally, we also confirm that a DQPT occurs for quenches in the opposite direction, BTC $\to$ non-BTC. Fig.~\ref{fig:FLE-quench-BTC-to-nBTC} shows $\mathcal{L}_F(t)$ for $N=100$ when quenching from $\omega_{0,i} = -1$ to  $\omega_{0,f} = 0$. Here $\mathcal{L}_F(t)$ drops below numerical precision at $\tco = 20.778$ and remains zero thereafter. Since when the post-quench Hamiltonian corresponds to non-BTC phase, the state of the system reaches a time-independent stationary state, vanishing of $\mathcal{L}_F$ for this late time stationary state with respect to the pre-quench initial state implies that the late-time stationary state becomes orthogonal to the initial state. Unlike the BTC case, here there is no late-time periodicity that could revive the overlap.

\begin{figure*}[t]
    \centering
    \begin{minipage}[t]{0.4\textwidth}
        \centering
        \includegraphics[trim=0cm 0.0cm 0.0cm 0, clip,width=7.5cm, height=5cm]{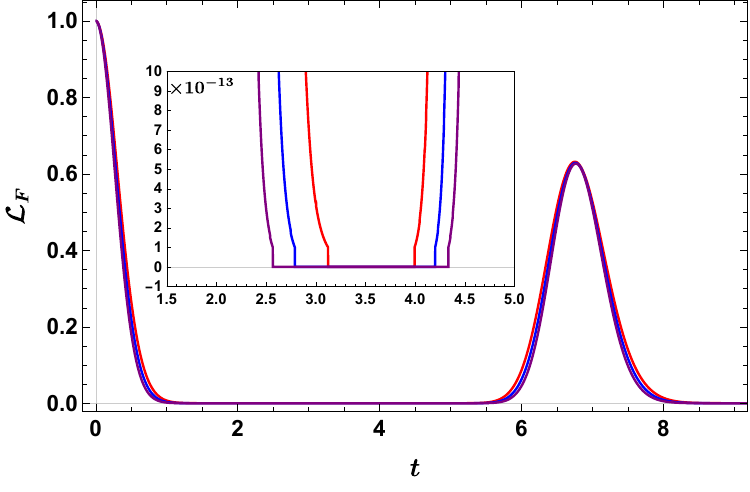}\\
        \textbf{(a)}
    \end{minipage}
    \hspace{1 cm}
    \begin{minipage}[t]{0.4\textwidth}
        \centering
        \includegraphics[trim=0cm 0.0cm 0.0cm 0, clip, width=7.5cm, height=5cm]{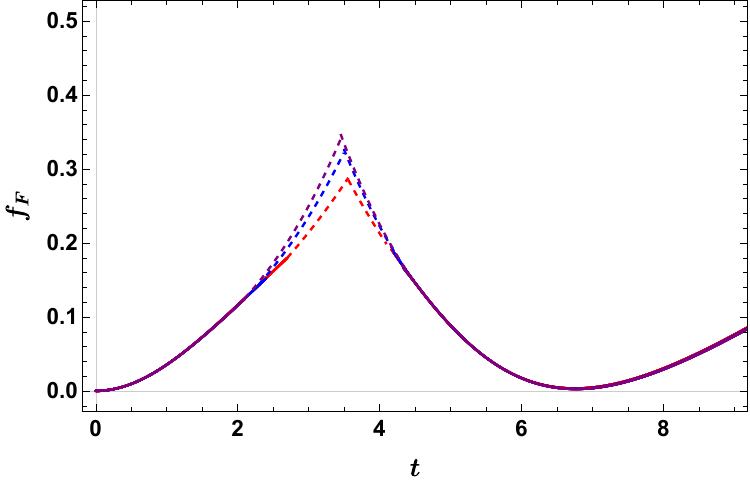}\\
        \textbf{(b)}
    \end{minipage}   \caption{\textit{DQPT under a finite-time ramp from the non-BTC to the BTC phase.} \textbf{(a)} The fidelity-based Loschmidt echo $\mathcal{L}_F$ is shown during the post-ramp unitary evolution for $N=70$ (red), $N=80$ (blue), and $N=90$ (purple). The parameter $\omega_0$ is ramped linearly from $\omega_{0,i} = 0$ to $\omega_{0,f} = -1$ over a duration $\tau = 5$ according to Eq.~\eqref{eq:ramp-omega0} under Lindblad dynamics, after which dissipation is switched off and the system evolves unitarily. $\mathcal{L}_F$ is computed with respect to the mixed state at the end of the ramp. The first zero of $\mathcal{L}_F$ defines the critical time $\tco$. The inset magnifies the region near the $t$-axis. Numerical zeros are identified using a threshold of $10^{-13}$. The values of $\tco$ are listed in Tab.~\ref{tab:FLE-ramp}. \textbf{(b)} The rate function $f_F$ corresponding to (a) is shown with the curves of the same colors for the respective $N$ values as in (a). 
    %for  $N=70$ (red), $N=80$ (blue), and $N=90$ (purple). 
    At the first critical time   $\tco$, where $\mathcal{L}_F$ vanishes, $f_F$ exhibits a cusp-like singularity, confirming the presence of a DQPT under the ramp protocol. Solid curves are generated for numerically accessible values, while dashed curves show extrapolations inside the numerically inaccessible region. }
    \label{fig:FLE-ramp}
\end{figure*}

\subsection{Ramp Protocol}
We now consider the finite-time ramp protocol described in Sec.~\ref{sec:setup::subsec:ramp}, where $\omega_0(t)$ is varied linearly from $\omega_{0,i} = 0$ to $\omega_{0,f} = -1$ over a duration $\tau=5$ as
\begin{equation}\label{eq:ramp-omega0}
    \omega_0(t) = \omega_{0,i} + (\omega_{0,f} - \omega_{0,i}) \frac{t}{\tau},
\end{equation}
with dissipation present during the ramp and switched off afterwards. We then compute $ \mathcal{L}_F(t')$ during the subsequent unitary evolution using $\rho(\tau)$ (end-of-ramp state) as the reference state. Specifically,
\begin{equation*}
    \mathcal{L}_F(t') = F(\rho(\tau), \rho(\tau + t')), \qquad t' \geq 0
\end{equation*}
as defined before in Eq.~\eqref{eq:LF-ramp}.

Fig.~\ref{fig:FLE-ramp}~(a) shows the plot of $\mathcal{L}_F$ for the ramp protocol with $N=70$ (red), $N=80$ (blue), and $N=90$ (purple). 
In contrast to the quench protocol, the reference state $\rho(\tau)$ in the ramp protocol is mixed, so $\mathcal{L}_F$ does not simplify to Eq.~\eqref{eq:FLE_pure_simplify} in this case. As a result, $\mathcal{L}_F$ must be evaluated using the Uhlmann fidelity [Eq.~\eqref{eq:Uhlmann-Fid}], which is numerically more demanding and more sensitive to accumulated floating-point noise when $\mathcal{L}_F$ becomes very small~\footnote{Note that Eq.~\eqref{eq:LF-ramp} involves products of large density matrices for larger $N$ values whereas Eq.~\eqref{eq:FLE_pure_simplify} computes just a single matrix element. The consequence of this is that the computation of Eq.~\eqref{eq:LF-ramp} is prone to faster accumulation of numerical noise during the time evolution compared to Eq.~\eqref{eq:FLE_pure_simplify}. This accumulation of noise during the time evolution is very prominent when the value of $\mathcal{L}_F$ becomes very small which happens precisely near its zeros.}. Subsequently, we are able to gather the data of $\mathcal{L}_F$ and corresponding $f_F$ only until $N=500$, and furthermore, we had to identify numerical zeros using the threshold $\varepsilon = 10^{-13}$, which is two orders of magnitude larger than the machine precision. As in the quench protocol, we define times $\tmo$ and $\tpo$ as the first times at which $\mathcal{L}_F$ respectively drops below and rises above the numerical zero. The first critical time is then obtained as $\tco = (\tmo + \tpo)/2$ as before. The resulting values are listed in Tab.~\ref{tab:FLE-ramp}. The corresponding rate function $f_F(t)$ is shown in Fig.~\ref{fig:FLE-ramp}~(b), which exhibits the expected singular behavior at $\tco$.

\begin{table}[t]
\centering
\setlength{\tabcolsep}{12pt}   % ← increase from default (~6pt)
\begin{tabular}{|c|c|c|c|}
\hline
\textbf{$N$} & \textbf{$\tmo$} & \textbf{$\tpo$} & \textbf{$\tco$} \\
\hline
 70 & 3.123 & 3.995 & 3.559 \\
\hline
 80 & 2.789 & 4.202 & 3.496 \\
\hline
 90 & 2.567 & 4.335 & 3.451 \\
\hline
\end{tabular}
\caption{\textit{Ramp protocol: non-BTC $\to$ BTC}. The values of $\tmo$, $\tpo$, and $\tco$ in Fig.~\ref{fig:FLE-ramp}~(a) corresponding to $N=70$, $80$, and $90$.}
\label{tab:FLE-ramp}
\end{table}

\begin{figure}[htbp]
    \centering
    \includegraphics[width=1\linewidth]{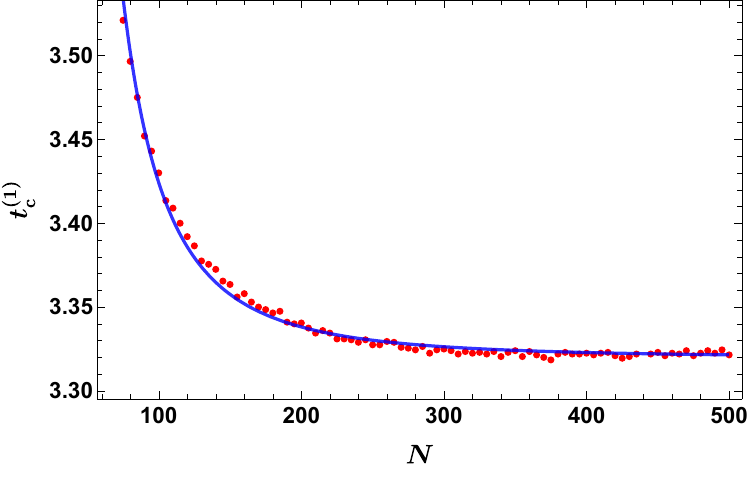}
    \caption{
    %\textbf{Ramp protocol: large-$N$ scaling of $\tco$} (non-BTC to BTC)- 
    \textit{Finite-size scaling of the first critical time for the ramp protocol.}
    The first critical time $\tco$ is plotted as a function of system size $N$ for the non-BTC $\to$ BTC ramp. The red dots represent the numerical data, while the blue curve is a power-law fit $\tco(N) = a N^{-b} + c$, demonstrating convergence to a constant value in the thermodynamic limit. The best-fit values of fitting parameters are $a = 1.045 \times 10^{4} \pm 9.513 \times 10^{1} $, $b=2.5$ and $c= 3.319 \pm (< 10^{-3})$. The RMS fitting error is $0.003$. The values of all the parameters are the same as in Fig.~\ref{fig:FLE-ramp}. 
      }
    \label{fig:tc-scaling-N-ramp}
\end{figure}

The finite-size scaling of the first critical time for the ramp protocol is shown in Fig.~\ref{fig:tc-scaling-N-ramp} for the same values of parameters used in Fig.~\ref{fig:FLE-ramp}. The data support convergence to a constant in the $N \to \infty$ limit, with a substantially faster approach than in the quench case for the accessible system sizes. The red dots are the actual data, whereas the blue curve is the best power-law fit of the form $t_c(N) = a N^{-b} + c$ for the full range of data shown in Fig.~\ref{fig:tc-scaling-N-ramp} with best-fit values of fitting parameters as $a = 1.045 \times 10^{4} \pm 9.513 \times 10^{1} $, $b=2.5$ and $c= 3.319 \pm (<10^{-3})$. The RMS fitting error is $0.003$. Since the time resolution used to generate the time evolution is $\Delta t = 0.001$, we truncate the best-fit parameter values and their errors at the third decimal place. Note that since the amount of data available for fitting is limited to $N \leq 500$ for the reasons described earlier the numerical routine employed for fitting is not able to fit the function $a N^{-b} + c$ with three parameters. To mitigate this, we have varied the parameter $b$ by hand while fitting the other two parameters for each choice of $b$ and found that the RMS error is minimized for $b=2.5$. In conclusion, we find that in the $N \to \infty$ limit $\tco$ approaches a value of 3.319 via a power-law with an exponent of $2.5$.

\section{Conclusion }\label{sec:conclusion}

{In this work, we have demonstrated that dynamical quantum phase transitions occur in a Markovian open quantum many-body system that supports a boundary time-crystal phase. Using the fidelity-based Loschmidt echo suitable for mixed-state Lindblad dynamics, we identified nonanalyticities in the corresponding rate function following both sudden quenches and finite-time ramps across the BTC transition. We extracted the finite-size scaling of the first critical time and showed convergence toward a well-defined thermodynamic-limit value, with distinct scaling exponents for quench and ramp protocols. When the post-quench parameters lie in the BTC phase, the system approaches a time-periodic steady state, and the Loschmidt echo exhibits repeated zeros, leading to a sequence of DQPT critical times. In contrast, when the dynamics relaxes to a stationary steady state outside the BTC phase, the first critical time can be followed by a regime without revivals. More broadly, our results show that DQPTs extend into genuinely dissipative time-crystalline phases and are not restricted to closed or Floquet systems. }

\acknowledgements
We acknowledge that the computations were performed using ARMADILLO~\cite{Sanderson2016, Sanderson_2018} on the cluster computing facility of the Harish-Chandra Research Institute, India.

% Appendices 
\appendix

\FloatBarrier

% References
%\clearpage
%\bibliographystyle{apsrev4-2}
%\bibliographystyle{unsrt}
\bibliography{references} % Add your .bib file here

% End document
\end{document}